\documentclass{article}

\usepackage{arxiv}

\usepackage{amsmath,amssymb,amsfonts}

\usepackage[utf8]{inputenc} 
\usepackage[T1]{fontenc}    
\usepackage{hyperref}       
\usepackage{url}            
\usepackage{booktabs}       
\usepackage{microtype}      
\usepackage{graphicx}
\usepackage{natbib}
\usepackage{doi}
\usepackage{textcomp}       
\usepackage{xcolor}

\usepackage{tikz}
\usetikzlibrary{automata,arrows,arrows.meta,positioning,calc,fit,backgrounds,decorations.pathreplacing}
\usepackage{subcaption}

\newcommand{\eg}{\emph{e.g.}}

\tikzset{
  box/.style={draw, rounded corners=2pt, minimum height=0.5cm, inner sep=2.5pt, align=center, font=\scriptsize\bfseries},
  encoderbox/.style={box, fill=orange!30, draw=orange!70!black},
  decoderbox/.style={box, fill=green!25, draw=green!55!black},
  normbox/.style={box, fill=yellow!40, draw=yellow!45!black},
  lbl/.style={align=center, inner sep=1pt},
  arr/.style={-{Latex[length=1.6mm]}, line width=0.7pt},
}
\newcommand{\drawSystemModel}[1]{%
  \begin{tikzpicture}[
      scale=#1, transform shape,
      font=\scriptsize,
      pics/ant/.style={code={
        \draw[black!85, line width=0.7pt] (0,-0.26) -- (0,0.10);
        \draw[black!85, line width=0.7pt] (-0.14,0.30) -- (0,0.10) -- (0.14,0.30);
        \draw[black!85, line width=0.7pt] (-0.16,-0.26) -- (0.16,-0.26);
      }},
      pics/rf/.style={code={
        \foreach \r in {0.28,0.42,0.56}
          \draw[blue!55!black, line width=0.6pt] (-38:\r) arc (-38:38:\r);
      }},
      pics/avatar/.style={code={
        \draw[draw=blue!55!black, fill=blue!12, rounded corners=1pt] (-0.21,0.30) rectangle (0.21,0.68);
        \draw[draw=blue!55!black, fill=white, rounded corners=0.5pt] (-0.15,0.35) rectangle (0.15,0.63);
        \fill[blue!55!black] (-0.055,0.545) circle (0.018);
        \fill[blue!55!black] (0.055,0.545) circle (0.018);
        \draw[blue!55!black, line width=0.35pt] (-0.05,0.44) .. controls (0,0.41) .. (0.05,0.44);
        \draw[draw=blue!55!black, fill=blue!25, rounded corners=1.5pt] (-0.27,-0.05) rectangle (0.27,0.28);
        \foreach \x in {-0.13,0,0.13} \fill[blue!60!black] (\x,0.19) circle (0.023);
        \draw[blue!55!black, line width=0.8pt, line cap=round] (-0.27,0.19) -- (-0.37,0.00);
        \draw[blue!55!black, line width=0.8pt, line cap=round] ( 0.27,0.19) -- ( 0.37,0.00);
        \draw[draw=blue!55!black, fill=blue!18] (-0.08,-0.40) rectangle (0.08,-0.05);
        \draw[draw=blue!55!black, fill=blue!25, rounded corners=1pt] (-0.27,-0.52) rectangle (0.27,-0.40);
        \fill[blue!55!black] (-0.17,-0.57) circle (0.048);
        \fill[blue!55!black] ( 0.17,-0.57) circle (0.048);
      }},
      pics/operator/.style={code={
        \draw[draw=blue!55!black, fill=blue!25]
          (-0.34,-0.52) .. controls (-0.34,-0.10) and (-0.15,0.02) .. (0,0.02)
          .. controls (0.15,0.02) and (0.34,-0.10) .. (0.34,-0.52) -- cycle;
        \draw[draw=blue!55!black, fill=blue!12] (0,0.26) circle (0.20);
        \draw[blue!55!black, line width=0.9pt] (-0.215,0.28) arc (180:0:0.215 and 0.235);
        \draw[draw=blue!55!black, fill=blue!60!black] (-0.245,0.17) rectangle (-0.185,0.30);
        \draw[blue!55!black, line width=0.6pt] (-0.215,0.17) .. controls (-0.20,0.05) and (-0.11,0.05) .. (-0.06,0.10);
      }},
    ]

    \pic at (-6.40, 0) {avatar};
    \node[lbl, anchor=north] at (-6.40, -0.64) {Cybernetic\\[-1pt] avatar};
    \node[encoderbox] (ienc) at (-4.80, 0) {Image\\[-1pt] Encoder $f_I$};
    \node[normbox] (ser) at (-2.50, 0) {Pair \&\\[-1pt] OFDM Mod.};

    \pic at (-0.85, 0) {ant};
    \pic at (-0.68, 0.10) {rf};
    \pic at (0.95, 0) {ant};

    \node[normbox] (rec) at (2.90, 0) {Equalize \&\\[-1pt] OFDM Demod.};
    \node[decoderbox] (cos) at (5.30, 0) {Cosine\\[-1pt] Similarity};
    \pic at (6.95, 0) {operator};
    \node[lbl, anchor=north] at (6.95, -0.64) {Operator};

    \node[encoderbox] (tenc) at (5.30, -1.10) {Text\\[-1pt] Encoder $f_T$};
    \node[lbl] (texts) at (2.90, -1.10) {Candidate texts\\[-1pt] $\{T_1,\ldots,T_K\}$};

    \draw[arr] (-5.97, 0) -- (ienc) node[midway, above, lbl] {$I$};
    \draw[arr] (ienc) -- (ser) node[midway, above, lbl] {$\mathbf{f}_I$};
    \draw[arr] (ser.east) -- (-1.12, 0) node[midway, above, lbl] {$\mathbf{z}$};
    \draw[arr] (1.18, 0) -- (rec.west) node[midway, above, lbl] {$\hat{\mathbf{z}}$};
    \draw[arr] (rec) -- (cos) node[midway, above, lbl] {$\hat{\mathbf{f}}_I$};
    \draw[arr] (cos) -- (6.58, 0) node[midway, above, lbl] {$\hat{T}$};
    \draw[arr] (texts) -- (tenc);
    \draw[arr] (tenc) -- (cos) node[midway, right, lbl] {$\mathbf{f}_{T_i}$};

    \node[lbl] at (-0.85, -0.46) {TX};
    \node[lbl] at (0.95, -0.46) {RX};
    \node[lbl] at (0.05, -0.80) {ADALM-Pluto transceivers};

  \end{tikzpicture}%
}

\tikzset{
  lbl/.style={align=center, inner sep=1pt},
  slot/.style={draw, minimum height=0.42cm, inner sep=1.5pt, align=center, font=\scriptsize},
  cpslot/.style={slot, fill=yellow!40, draw=yellow!45!black},
  symslot/.style={slot, fill=orange!30, draw=orange!70!black},
  binpilot/.style={draw=blue!55!black, fill=blue!30, line width=0.4pt},
  bindata/.style={draw=orange!70!black, fill=orange!30, line width=0.4pt},
  binfree/.style={draw=black!35, fill=black!6, line width=0.4pt},
}
\newcommand{\drawFrameStructure}[1]{%
  \begin{tikzpicture}[scale=#1, transform shape, font=\scriptsize]

    \node[lbl, anchor=east] at (-0.15, 0) {Time};
    \node[cpslot,  minimum width=1.01cm] (cp)   at (0.505, 0) {CP\\[-1pt] 64};
    \node[symslot, minimum width=4.04cm] (sym)  at (3.030, 0) {OFDM symbol, $N{=}512$};

    \draw[decorate, decoration={brace, amplitude=3pt, mirror}, line width=0.5pt]
      (0.0, -0.30) -- (5.05, -0.30)
      node[midway, below, lbl, yshift=-2pt] {576 samples at 2\,MHz = 288\,\textmu s};

    \node[lbl, anchor=east] at (-0.15, -1.55) {Freq.};
    \foreach \i in {0,...,31} {
      \pgfmathsetmacro{\xpos}{0.02 + \i*0.158}
      \pgfmathparse{int(mod(\i,4))}
      \ifnum\pgfmathresult=0
        \node[binpilot, minimum width=0.13cm, minimum height=0.36cm, inner sep=0pt]
          at (\xpos, -1.55) {};
      \else
        \ifnum\i<22
          \node[bindata, minimum width=0.13cm, minimum height=0.36cm, inner sep=0pt]
            at (\xpos, -1.55) {};
        \else
          \node[binfree, minimum width=0.13cm, minimum height=0.36cm, inner sep=0pt]
            at (\xpos, -1.55) {};
        \fi
      \fi
    }

    \draw[decorate, decoration={brace, amplitude=3pt}, line width=0.5pt]
      (-0.05, -1.30) -- (3.42, -1.30)
      node[midway, above, lbl, yshift=2pt] {256 data subcarriers};
    \draw[decorate, decoration={brace, amplitude=3pt}, line width=0.5pt]
      (3.50, -1.30) -- (5.02, -1.30)
      node[midway, above, lbl, yshift=2pt] {116 unused};

  \end{tikzpicture}%
}

\tikzset{
  hlbl/.style={align=center, inner sep=1pt, font=\footnotesize},
  hsml/.style={align=center, inner sep=1pt, font=\scriptsize},
  usbwire/.style={line width=0.9pt, draw=black!70, rounded corners=3pt},
  coax/.style={line width=1.2pt, draw=blue!55!black, rounded corners=4pt},
}
\def\hwicon{0.8}
\newcommand{\drawHardwareSetup}[1]{%
  \begin{tikzpicture}[
      scale=#1, transform shape, font=\footnotesize,
      pics/pluto/.style={code={
        \draw[draw=teal!65!black, fill=teal!28, rounded corners=1.5pt]
          (-1.00,-0.40) rectangle (1.00,0.40);
        \draw[draw=teal!65!black, fill=teal!10, rounded corners=1pt]
          (-0.86,-0.22) rectangle (0.62,0.22);
        \foreach \y in {0.17,-0.17}
          \draw[draw=yellow!35!black, fill=yellow!70!black!55]
            (1.00,\y-0.05) rectangle (1.15,\y+0.05);
        \draw[draw=black!60, fill=black!30] (-1.13,-0.07) rectangle (-1.00,0.07);
        \fill[green!60!black] (0.80,0.30) circle (0.04);
      }},
      pics/pi/.style={code={
        \draw[draw=green!45!black, fill=green!55!black!18, rounded corners=2pt]
          (-1.50,-0.58) rectangle (1.50,0.58);
        \draw[draw=black!70, fill=black!72] (-1.26,0.36) rectangle (1.12,0.52);
        \foreach \i in {0,...,18}
          \draw[black!35, line width=0.25pt]
            (-1.20+\i*0.125,0.38) -- (-1.20+\i*0.125,0.50);
        \draw[draw=black!55, fill=black!45] (-0.28,-0.12) rectangle (0.14,0.22);
        \draw[draw=black!45, fill=black!22] (1.06,-0.52) rectangle (1.56,-0.28);
        \draw[draw=black!45, fill=black!22] (1.06,-0.22) rectangle (1.56,0.02);
        \draw[draw=black!45, fill=black!30] (1.06,0.08) rectangle (1.56,0.32);
        \foreach \x in {-1.30,-0.84,-0.46}
          \draw[draw=black!60, fill=black!35] (\x,-0.66) rectangle (\x+0.18,-0.58);
        \foreach \p in {(-1.38,0.46),(1.24,0.46),(-1.38,-0.46),(1.24,-0.46)}
          \draw[black!40, fill=white] \p circle (0.042);
      }},
    ]

    \pic[scale=\hwicon] at (-2.30, 0) {pi};
    \node[hlbl, anchor=south] at (-2.30, 0.52) {Raspberry Pi~4 / Pi~5};

    \pic[scale=\hwicon] at (0.45, 0) {pluto};
    \pic[xscale=-\hwicon, yscale=\hwicon] at (2.79, 0) {pluto};
    \node[hsml] at (0.35, 0) {\textsc{tx}};
    \node[hsml] at (2.89, 0) {\textsc{rx}};
    \node[hlbl, anchor=south] at (0.45, 0.36) {ADALM-Pluto};
    \node[hlbl, anchor=south] at (2.79, 0.36) {ADALM-Pluto};

    \draw[coax] (1.37, 0.136) -- (1.87, 0.136);
    \node[hsml, anchor=north] at (1.62, -0.36) {coaxial cable};

    \draw[usbwire] (-1.05, -0.02) -- (-0.455, -0.02);
    \draw[usbwire] (3.70, 0) -- (3.95, 0) -- (3.95, -0.80) -- (-0.85, -0.80)
      -- (-0.85, -0.32) -- (-1.05, -0.32);
    \node[hsml, anchor=south] at (-0.75, 0.03) {USB};
    \node[hsml, anchor=south] at (3.40, -0.80) {USB};

  \end{tikzpicture}%
}

\title{Implementation of Zero-shot Semantic Communication on Software Defined Radio}

\author{Thanh Le \quad Arif Dataesatu \quad Homare Murakami \quad Takeshi Matsumura \\
  Wireless Systems Laboratory, Wireless Networks Research Center \\
  National Institute of Information and Communications Technology (NICT) \\
  Yokosuka, Kanagawa, Japan}

\date{}

\renewcommand{\undertitle}{}
\renewcommand{\shorttitle}{Zero-shot Semantic Communication on Software Defined Radio}

\hypersetup{
  pdftitle={Implementation of Zero-shot Semantic Communication on Software Defined Radio},
  pdfauthor={Thanh Le, Arif Dataesatu, Homare Murakami, Takeshi Matsumura},
  pdfkeywords={orthogonal frequency-division multiplexing, software-defined radio, vision-language model, zero-shot classification},
}

\begin{document}
\maketitle

\begin{abstract}
    Semantic communication has recently gained traction for its ability to reduce the amount of data transmitted over a communication link
    by transmitting a task-oriented representation instead of the raw source.
    Zero-shot semantic communication sends a general embedding from a vision-language model (VLM),
    so the same transmitter can serve new classification tasks without retraining.
    Most evidence for this advantage, however, comes from numerical simulation.
    We implement zero-shot semantic communication on a software-defined radio platform:
    a Raspberry Pi drives a pair of Analog Devices Active Learning Module (ADALM)-Pluto transceivers,
    with an image encoder at the transmitter and a text encoder at the receiver,
    and determines the zero-shot classification results via cosine similarity.
    We compare two VLMs, CLIP and MobileCLIP,
    across various channel conditions, \emph{i.e.}, different signal-to-noise ratios (SNRs).
    We validate that the semantic link spends $9\times$ fewer channel uses per image than a JPEG plus 16-ary quadrature amplitude modulation baseline
    and still reaches 82\% accuracy on CIFAR-10 at $22.3$\,dB, where the baseline scores 0\%.
    On the traffic sign recognition dataset (TSRD), MobileCLIP correctly classifies 98.3\% of unseen images at the same SNR.
    Offloading the image encoder to a neural processing unit reduces encoding to $13.4$\,ms per image, $49\times$ faster than a Raspberry Pi~4 CPU, placing the transmitter within a real-time budget.
    Our implementation is publicly available at \url{https://github.com/thanhlexyz/zsscsdr}.
\end{abstract}


\section{Introduction}
    \label{sec:introduction}

    Semantic communication transmits a compact representation to the receiver,
    optimized for the specific task~\citep{le2026verifying}.
    For instance, a receiver tasked with recognizing a traffic sign requires data that conveys the sign's meaning
    rather than the original high-resolution image.
    Conventional wireless communication systems use bandwidth to reconstruct the entire image,
    even though the receiver ultimately discards most of it.
    Task-oriented semantic communication implements semantic communication
    by encoding only the information necessary for a given task~\citep{xi2021taskoriented,liu2025goaloriented}.
    However, this approach is limited because the encoder is trained for a fixed task and dataset,
    necessitating retraining and redeployment when encountering previously unseen data.

    Foundation models, including vision-language models (VLMs)~\citep{radford2021learning,vasu2024mobileclip},
    address this challenge by having general embedding spaces that are inherently task-agnostic.
    Recent studies have transmitted such embeddings for zero-shot inference,
    which involves classifying into any user-defined labels without task-specific training~\citep{hu2025zeroshot}.
    However, these results are primarily based on numerical simulations.
    Simulations do not account for nonidealities such as imperfect channel estimates.
    Therefore, this work seeks to determine whether the advantages of zero-shot semantic communication persist
    when implemented on a software-defined radio (SDR).
    Previous research has validated learned deep joint source-channel coding~\citep{bourtsoulatze2019deep},
    a form of semantic communication that still attempts to recover the full data,
    on SDR~\citep{ding2026prototyping},
    but has not addressed the zero-shot variant.

    Cybernetic avatar systems are a natural application for zero-shot semantic communication.
    A cybernetic avatar is a robotic system that acts as a teleoperator's surrogate in a remote environment,
    over a Local~5G network shared with other avatars~\citep{murakami2025towards,murakami2026toward}.
    Future deployments envision many teleoperators collaboratively controlling many avatars~\citep{hagita2024cybernetic},
    so perception streams, control commands, and safety traffic compete for a finite radio pool~\citep{dataesatu2024multilayer,dataesatu2025fair}.
    Perception streams are the bulk of that load and the part least sensitive to reconstruction,
    so compressing them semantically frees bandwidth for the traffic that cannot wait~\citep{ekolle2025agoal,ekolle2026asemantic}.
    Because avatars handle general tasks that cannot all be specified at deployment,
    a zero-shot link that needs no retraining for new categories fits the operational model.
    \begin{figure}[tbp]
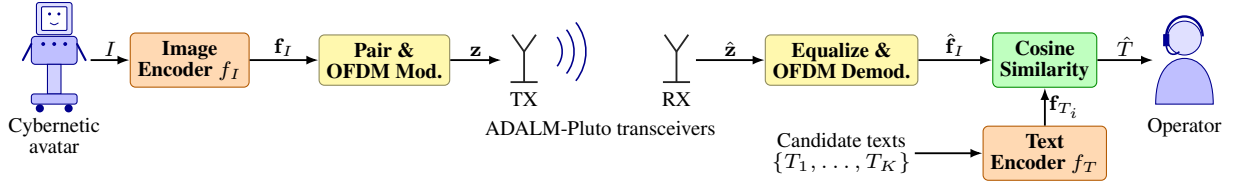

        \centering
        \drawSystemModel{1.12}
        \caption{Zero-shot semantic communication system model for transmitting cybernetic avatars' observations to teleoperators.}
        \label{fig:system-model}
    \end{figure}

    To our knowledge, zero-shot semantic communication has not yet been demonstrated using radio hardware.
    We address this gap through the following contributions:
        \begin{itemize}
            \item \textbf{Hardware implementation.}
                A Raspberry Pi operates a pair of Analog Devices Active Learning Module (ADALM)-Pluto transceivers.
                A VLM encoder generates the payload, and the receiver selects outputs by cosine similarity
                against text prompts that can be updated at any time without retraining.
            \item \textbf{VLM comparison.}
                We compare two VLM encoders,
                Contrastive Language-Image Pre-training (CLIP) and the mobile-scale MobileCLIP,
                quantifying the accuracy each achieves relative to its encoding time on the Raspberry Pi.
                Offloading the encoder to a neural processing unit (NPU) reduces encoding to $13.4$\,ms per image, $49\times$ faster than a Raspberry Pi~4 CPU.
            \item \textbf{Measured link performance.}
                Under varying transmit power, the semantic link needs $9\times$ fewer channel uses than the conventional baseline
                and holds higher accuracy under noise: 98.3\% on the traffic sign recognition dataset (TSRD) at 22.3\,dB with MobileCLIP,
                where the baseline scores 0\%.
        \end{itemize}

\section{Overview}
    \label{sec:model}

    \subsection{Vision-Language Model}
        \label{sec:model:vlm}
        Conventional vision models are limited to a predefined set of labels.
        Classifying new data requires collecting new labels and retraining the model.
        CLIP~\citep{radford2021learning}
        and its variants~\citep{vasu2024mobileclip} utilize natural language captions for supervision,
        pairing an image encoder $f_I(\cdot)$ with a text encoder $f_T(\cdot)$ in a shared $d$-dimensional space.
        The model is trained on millions of image-text pairs to handle a wide variety of topics.
        During training on a batch of $M$ pairs,
        the model learns which of the $M \times M$ pairings are correct
        by increasing the cosine similarity of the $M$ true pairs
        and decreasing that of the remaining $M^{2}-M$ incorrect pairs.
        Let $l \triangleq [0,1,\ldots,M-1]$ denote the correct indices.
        With $\mathbf{f}_I,\mathbf{f}_T$ stacking the $M$ batch embeddings, the loss is a symmetric cross-entropy over their similarity matrix.
        \begin{equation}
            \label{eq:clip}
            \mathcal{L}_{\mathrm{CLIP}} = \Bigl[\mathcal{L}_{ce}\bigl(\cos(\mathbf{f}_I,\mathbf{f}_T^{\top}), l\bigr) + \mathcal{L}_{ce}\bigl(\cos(\mathbf{f}_T,\mathbf{f}_I^{\top}), l\bigr)\Bigr]/2 .
        \end{equation}
        This loss function consists of two cross-entropy losses: the first asks which text matches each image, while the second asks which image matches each text.

        A trained CLIP model can recognize previously unseen class sets by representing each class as a text caption,
        encoding it with the text encoder $f_T(\cdot)$, and assigning images to the nearest text embedding.
        Task adaptation is achieved by modifying the caption rather than retraining the classification model,
        a design choice that supports general data transmission.

    \subsection{System Model}
        \label{sec:model:system}
        We consider a remote monitoring task within a cyber-physical system,
        \eg, a cybernetic avatar transmitting image observations to a teleoperator via a wireless link.
        \autoref{fig:system-model} illustrates the complete end-to-end pipeline.
        The cybernetic avatar captures an image $I$,
        which is processed by a pretrained image encoder to produce the embedding $\mathbf{f}_I = f_I(I)$.
        The transmitter pairs the $d$ real components of $\mathbf{f}_I$ into $L = d/2$ complex symbols
        and modulates them onto orthogonal frequency-division multiplexing (OFDM) subcarriers as $\mathbf{z} \in \mathbb{C}^{L}$,
        subject to the average power constraint $\lVert\mathbf{z}\rVert^{2}/L \le P$.

        The symbols are transmitted to the receiver over a block Rayleigh fading channel.
        \begin{equation}
            \label{eq:fading}
            \hat{\mathbf{z}} = h\,\mathbf{z} + \mathbf{n}.
        \end{equation}
        In this model, the gain $h$ is assumed to be distributed as $\mathcal{CN}(0,1)$, remaining constant during the transmission of a single image and varying independently between images.
        The noise $\mathbf{n} \in \mathbb{C}^{L}$ is modeled as white Gaussian noise, with each entry distributed as $n_i \sim \mathcal{CN}(0,\sigma^{2})$.
        Equation~\eqref{eq:fading} represents the standard assumption for the wireless link.
        However, in this implementation, neither $h$ nor $\mathbf{n}$ is sampled,
        as both are generated by the physical link between the two SDR transceivers.
        The signal-to-noise ratio (SNR) of the wireless link is defined as $\mu = 10\log_{10}(P/\sigma^{2})$\,dB.

        The receiver equalizes and demodulates the channel output $\hat{\mathbf{z}}$ to obtain a noisy embedding $\hat{\mathbf{f}}_I$.
        The receiver maintains a set of candidate texts $\{T_1,\ldots,T_K\}$ corresponding to the classes to be distinguished.
        Each candidate text is encoded using the pretrained text encoder as $\mathbf{f}_{T_i} = f_T(T_i)$.
        The receiver then selects the label whose text embedding is closest to the received image embedding.
        \begin{equation}
            \label{eq:decision}
            \hat{T} = \arg\max_{T_i} \cos\bigl(\hat{\mathbf{f}}_I, \mathbf{f}_{T_i}\bigr).
        \end{equation}
        The selected label $\hat{T}$ is then communicated to the operator.
        Both encoders are pretrained and operate in inference mode only, without retraining;
        hence, serving a new class list requires only updating the candidate texts at the receiver.

\section{Proposed Implementation}
    \label{sec:implementation}

    \subsection{Hardware and Software Architecture}
        \label{sec:implementation:hardware}
\begin{figure}[tbp]
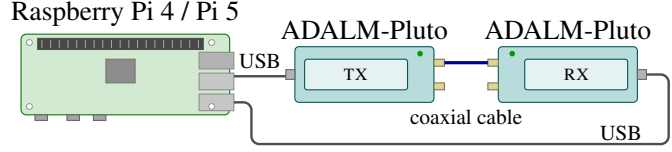

            \centering
            \drawHardwareSetup{1.15}
            \caption{The SDR hardware and its connections.}
            \label{fig:hardware}
        \end{figure}

        A Raspberry Pi~4 or Pi~5 hosts the demonstration and controls
        two ADALM-Pluto transceivers via USB (\autoref{fig:hardware}).
        The two SubMiniature version A (SMA) ports on the two ADALM-Pluto transceivers are directly connected by a coaxial cable.
        The image encoder $f_I$ and the OFDM modulator are implemented on the transmitter,
        whereas equalization, demodulation, the text encoder $f_T$, and cosine similarity computation are performed on the receiver.
        The link operates at a carrier frequency of 915\,MHz, well inside the ADALM-Pluto tuning range of 325\,MHz to 3.8\,GHz.
        The 915\,MHz band carries LoRa and other industrial Internet-of-Things traffic,
        so it is representative of the unlicensed sub-GHz spectrum available to drones, autonomous robots, or cybernetic avatars.

        We deliberately designed this architecture to maintain a modest hardware footprint.
        The Raspberry Pi consumes only a few watts, \eg, less than $7$\,W for the Pi~4,
        which is within the operational limits of drones and mobile robots,
        so it serves as a representative computing platform for these applications.
        All other components are commodity and open-source,
        including off-the-shelf ADALM-Pluto transceivers and the \texttt{pyadi-iio}, CLIP, and OpenCLIP Python libraries.
        The Python implementation is available at \url{https://github.com/thanhlexyz/zsscsdr}.
        This availability enables cost-effective reproduction and facilitates further development based on the proposed implementation.

    \subsection{OFDM Implementation}
        \label{sec:implementation:ofdm}
        OFDM serves as the data transmission scheme responsible for carrying the embedding over the radio link~\citep{wu2022channeladaptive}.
        OFDM also forms the foundation of the physical layer in 4G, 5G, and recent Wi-Fi standards.
        Consequently, a demonstration based on OFDM can be readily transferred to deployed systems.
        The structural elements common to these standards, such as the cyclic prefix and comb-type pilots, are retained.
        The SDR demonstration is implemented from scratch because full protocol stacks are large
        and do not build efficiently on the Raspberry Pi and ADALM-Pluto devices.
        Our minimal implementation enabled a fast and compatible prototype on the chosen hardware.

\begin{figure}[tbp]
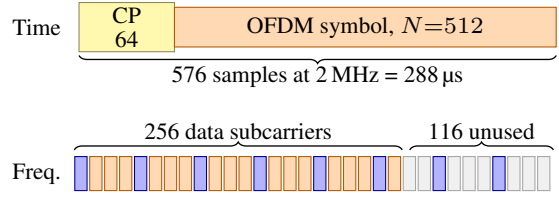

            \centering
            \drawFrameStructure{1.25}
            \caption{OFDM frame structure; blue cells mark the comb pilots.}
            \label{fig:frame-structure}
        \end{figure}

        \autoref{fig:frame-structure} illustrates the frame structure, which consists of a 64-sample cyclic prefix followed by a single 512-point OFDM symbol, totaling 576 samples or 288\,\textmu s at 2\,MHz.
        Each embedding is transmitted within a single OFDM frame, with the cyclic prefix configured to mitigate multipath propagation.
        In the absence of a guard interval, multipath propagation causes symbol overlap, resulting in inter-symbol interference and a frequency-selective channel.
        Prepending a copy of the symbol tail to the output of the inverse Fourier transform renders the channel convolution circular.
        The forward transform subsequently diagonalizes the channel into orthogonal subchannels, so that~\eqref{eq:fading} holds on each subcarrier with its own coefficient~$h$.
        The receiver recovers these coefficients per frame from the pilots, after removing the carrier frequency offset estimated from the cyclic prefix and pilots.
        The 64-sample cyclic prefix, one eighth of the symbol or 32\,\textmu s, exceeds the typical indoor delay spread at 915\,MHz.

        The ADALM-Pluto is a direct-conversion transceiver,
        so local-oscillator leakage and offsets in the in-phase and quadrature signal paths manifest as a spurious tone.
        This spurious tone appears on the subcarrier located at the center of the frequency band.
        Conversely, the analog filter exhibits roll-off near half the sample rate, attenuating the subcarriers at the edges of the frequency band.
        As a result, nine subcarriers centered on each of these two points are reserved and not used for data transmission.
        Six of these 18 coincide with the 128 comb pilots on every fourth subcarrier, so twelve are nulled and $512-128-12=372$ remain available.
        One embedding occupies the first 256 of the 372, and the remaining 116 are left unused.
        The $\ell_2$-normalized embedding meets the constraint $\lVert\mathbf{z}\rVert^{2}/L \le P$ of \autoref{sec:model:system}, and the transmitter scales each frame to the same peak amplitude.

        The receiver does not perform closed-loop tracking; consequently, each frame arrives at an unknown offset within the receive buffer.
        Frame timing is the peak of the cyclic-prefix correlation at a 512-sample lag, accumulated over the cyclically repeated frames in the receive buffer.
        \autoref{tab:ofdm} summarizes the resulting parameters and the transceiver settings.

        \begin{table}[tbp]
            \centering
            \caption{OFDM and transceiver parameters}
            \label{tab:ofdm}
            \begin{tabular}{@{}ll@{}}
                \toprule
                Parameter & Value \\
                \midrule
                Transceivers & 2 $\times$ ADALM-Pluto \\
                Host & Pi~4 (link), Pi~5 (processing time) \\
                Carrier frequency & 915\,MHz \\
                Sample rate & 2\,MHz \\
                FFT size $N$ & 512 \\
                Subcarrier spacing & 3.91\,kHz \\
                Cyclic prefix & 64 samples (32\,\textmu s) \\
                Pilot spacing & every 4th subcarrier (128 pilots) \\
                Data subcarriers & 256 of 372 usable subcarriers \\
                Frame length & 576 samples (288\,\textmu s) \\
                Nominal TX power & $-73$ to $-13$\,dBm (swept) \\
                RX gain & $30$\,dB, manual \\
                Measured SNR range & $-2.6$ to $43.7$\,dB \\
                \bottomrule
            \end{tabular}
        \end{table}

\section{Experiment}
    \label{sec:experiment}

    Our evaluation answers the following research questions:
    \noindent\textbf{RQ1}: How does zero-shot accuracy vary across channel conditions and unseen datasets? \\
    \noindent\textbf{RQ2}: How much bandwidth does zero-shot semantic transmission save compared to a traditional baseline? \\
    \noindent\textbf{RQ3}: What is the processing time on the Raspberry Pi?

    \begin{figure}[tbp]
        \centering
        \includegraphics[width=0.72\linewidth]{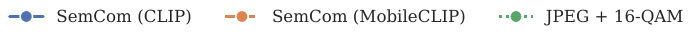}
        \\[0.4em]
        \begin{subfigure}[b]{0.48\linewidth}
            \centering
            \includegraphics[width=\linewidth]{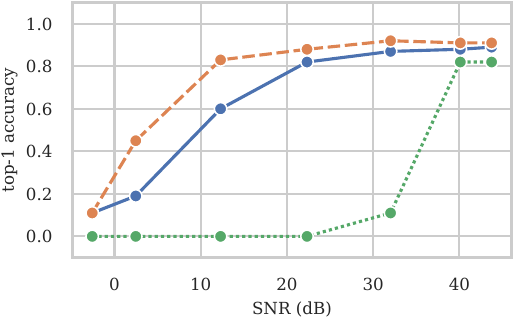}
            \caption{CIFAR-10}
            \label{fig:rq1-cifar10}
        \end{subfigure}
        \hfill
        \begin{subfigure}[b]{0.48\linewidth}
            \centering
            \includegraphics[width=\linewidth]{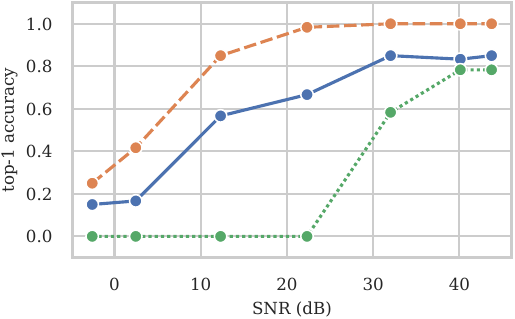}
            \caption{TSRD}
            \label{fig:rq1-tsrd}
        \end{subfigure}
        \caption{Accuracy of zero-shot semantic communication (SemCom) and the JPEG + 16-QAM baseline versus measured SNR.}
        \label{fig:rq1}
    \end{figure}

    \subsection{Experimental Setup}
        \label{sec:experiment:setup}

        \noindent\textbf{Platform.}
            Link measurements use a Raspberry Pi~4; processing time is additionally reported on a Raspberry Pi~5, on its CPU and with a Hailo-10H NPU over PCIe.

        \noindent\textbf{Model.}
            This study evaluates two VLMs, both pretrained and used exclusively for inference in their released forms: CLIP ViT-B/32~\citep{radford2021learning} and MobileCLIP2-S0 from the MobileCLIP family~\citep{vasu2024mobileclip}, hereafter CLIP and MobileCLIP.
            Both models use a $d{=}512$ embedding space, so each image feature requires $L{=}256$ complex symbols, and the OFDM configurations are identical.
            Image embeddings undergo $\ell_2$ normalization before transmission.

\begin{figure}[tbp]
                \centering
                \includegraphics[width=0.47\linewidth]{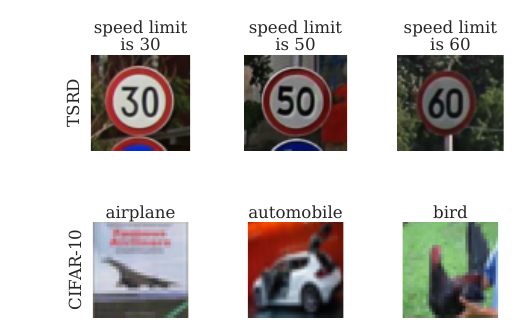}
                \caption{Three example classes from each dataset (first evaluated image of each class): TSRD signs at native resolution and CIFAR-10 thumbnails at $32{\times}32$.}
                \label{fig:dataset-examples}
            \end{figure}

        \noindent\textbf{Datasets and prompts.}
            We evaluate on two datasets that differ in resolution, namely, CIFAR-10 and TSRD.
            CIFAR-10 provides $32{\times}32$ images across ten classes.
            We use one hundred test images and the prompt template ``a photo of a \{class\}''.
            TSRD~\citep{yang2016towards} provides traffic sign images of varying size.
            We use 60 images (median $112{\times}106$ pixels) from six of its classes, 10 per class.
            The candidate prompts for this dataset are:
            ``speed limit is 30'', ``speed limit is 50'', ``speed limit is 60'', ``speed limit is 80'', ``pedestrian crossing'', and ``do not enter''.
            For CIFAR-10, we use the first ten test indices of each class, resulting in a fixed image set rather than a randomly sampled one.
            \autoref{fig:dataset-examples} shows three example classes from each dataset.

        \noindent\textbf{Baseline.}
            The conventional transmission pipeline transmits the original image data directly.
            JPEG compression is applied to each image, with the quality parameter determined by a binary search to meet a 1146-byte size constraint
            and optimized Huffman tables. If the lowest quality setting exceeds this limit, the system incrementally reduces image resolution.
            We tuned this budget to the smallest size at which the baseline still reaches roughly $80\%$ accuracy on a clean channel (\autoref{sec:experiment:snr});
            a tighter budget would starve the baseline regardless of channel condition, thereby weakening the comparison.

            The resulting bitstream includes a two-byte synchronization word, a two-byte length field, and a 16-bit cyclic redundancy check (CRC) for error detection.
            With 16-ary quadrature amplitude modulation (16-QAM), the compressed image fills nine OFDM frames.
            At the receiver, the image is decoded and classified using the same CLIP model~\citep{radford2021learning}.

        \noindent\textbf{Channel.}
            All results are obtained using the link described in \autoref{sec:implementation:hardware},
            employing the OFDM parameters specified in \autoref{tab:ofdm}.
            Transmit power determines the operating point and is the sole parameter varied in these experiments.
            The ADALM-Pluto transceiver has a maximum transmission power of $+7$\,dBm.
            The nominal transmit power is swept across $\{-73,-63,\ldots,-13\}$\,dBm.
            Receiver gain remains fixed in manual mode;
            therefore, any change in transmit power is reflected in the SNR of the channel.

            At each transmit power setting, 200 probe frames consisting of unit-power complex Gaussian tones are used to determine the operating SNR,
            calculated from the residual error power after removal of the common phase.
            The seven transmit power settings correspond to measured SNRs of
            $-2.6$, $2.5$, $12.3$, $22.3$, $32.0$, $40.1$, and $43.7$\,dB.
            We plot all subsequent accuracy results against these measured SNR values.

        \noindent\textbf{Metrics.}
            We report accuracy as the proportion of images for which the nearest candidate prompt corresponds to the ground-truth class.
            We measure bandwidth in channel uses per image because the semantic path transmits analog symbols and does not generate a measurable bitstream.
            For RQ3, we report the median wall-clock time for both transmit-side and receive-side processing, excluding OFDM, 16-QAM, and the transceivers.

    \subsection{Accuracy Across Channel Conditions}
        \label{sec:experiment:snr}
            When evaluated against the link with different measured SNRs,
            the semantic and conventional accuracies exhibit distinct trends (\autoref{fig:rq1}).
            As SNR decreases, both zero-shot semantic transmission paths gradually decline in performance, ultimately approaching chance level.
            For CLIP, CIFAR-10 classification accuracy decreases from 89\% to 11\% as the link SNR degrades from $43.7$ to $-2.6$\,dB.

            On TSRD, zero-shot semantic communication demonstrates a similar trend,
            with CLIP accuracy decreasing from 85\% to 15\%, compared to a 16.7\% accuracy of a random classifier for six classes.
            Notably, this behavior is independent of the specific model used to generate the embedding, \eg, both CLIP and MobileCLIP have relatively similar trends.

            In contrast, the baseline accuracy remains at exactly 0\% for all SNR values at or below 22.3\,dB across both datasets.
            The minimum accuracy is zero because frames that fail synchronization or the CRC produce no output image.
            In such cases, there is no data to decode or re-embed, and the trial is consequently recorded as incorrect.
            Above this threshold, the baseline accuracy increases sharply, reaching 11\% at 32.0\,dB and 82\% at 40.1\,dB on CIFAR-10.

            Zero-shot semantic communication and the baseline have a substantial performance gap at low SNR.
            With CLIP, semantic transmission achieves 82\% accuracy on CIFAR-10 at 22.3\,dB.
            The baseline does not attain comparable accuracy until 40.1\,dB.
            Across the four SNR settings from $-2.6$ to $22.3$\,dB, the baseline fails to classify any samples.
            Once the link quality improves, the gap narrows, with accuracies of 89\% and 82\% on CIFAR-10,
            indicating that the primary benefit of zero-shot semantic communication is concentrated in the low-SNR regime.

            MobileCLIP matches or exceeds CLIP's semantic transmission accuracy at every SNR on both datasets,
            with the largest gains between 2.5 and 22.3\,dB.
            On CIFAR-10, MobileCLIP reads 45\% against CLIP's 19\% at 2.5\,dB and 83\% against 60\% at 12.3\,dB.
            In the best channel condition, the two achieve 91\% and 89\%.
            TSRD shows the same pattern more sharply, with gains of 10.0 to 31.7 percentage points.
            \begin{figure}[tbp]
                \centering
                \includegraphics[width=0.54\linewidth]{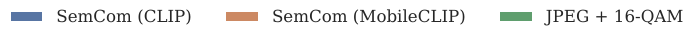}
                \\[0.4em]
                \begin{subfigure}[b]{0.26\linewidth}
                    \centering
                    \includegraphics[width=\linewidth]{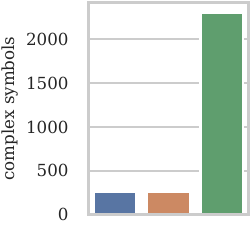}
                    \caption{Channel uses}
                    \label{fig:rq2-payload}
                \end{subfigure}
                \hspace{1.5em}
                \begin{subfigure}[b]{0.26\linewidth}
                    \centering
                    \includegraphics[width=\linewidth]{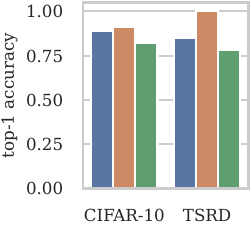}
                    \caption{Accuracy at $43.7$\,dB}
                    \label{fig:rq2-accuracy}
                \end{subfigure}
                \caption{Cost versus accuracy. Zero-shot semantic communication needs $9\times$ fewer channel uses and is more accurate.}
                \label{fig:rq2}
            \end{figure}

    \subsection{Bandwidth and Cross-Dataset Comparison}
        \label{sec:experiment:payload}
        The semantic payload requires a single OFDM frame with 256 tones (\autoref{fig:rq2}).
        The baseline approach requires nine frames, totaling 2304 tones, because 1146 JPEG bytes encoded at four bits per tone cannot fit in fewer frames.
        Consequently, semantic transmission needs $9\times$ fewer channel uses per image.

        Furthermore, semantic transmission achieves higher classification accuracy across both evaluated models.
        At the highest evaluated setting, CLIP achieves 89\% accuracy on CIFAR-10 versus the baseline's 82\%, and 85.0\% versus 78.3\% on TSRD.
        MobileCLIP achieves 91\% accuracy on CIFAR-10 and 100\% on TSRD.
        The more efficient representation yields higher accuracy because the baseline discards task-relevant details to meet its bit budget, resulting in semantic recovery from a degraded reconstruction.

        For CLIP, the performance margin remains consistent across both datasets, despite a median $12\times$ difference in pixel count, with improvements of 7.0 and 6.7 percentage points.
        Although a TSRD sign provides JPEG with substantially more data to compress than a $32{\times}32$ thumbnail, the semantic advantage remains nearly unchanged.
        Therefore, the observed advantage is attributable to the representation itself.
        MobileCLIP increases the margin to 9.0 and 21.7 percentage points, indicating that model selection affects the magnitude of the advantage but not its direction.

        \begin{table}[tbp]
            \centering
            \caption{Median per-image encode and decode time across compute platforms, in milliseconds}
            \label{tab:runtime}
            \setlength{\tabcolsep}{3pt}
            \begin{tabular}{@{}llrrrr@{}}
                \toprule
                & & \multicolumn{2}{c}{CIFAR-10} & \multicolumn{2}{c}{TSRD} \\
                \cmidrule(lr){3-4}\cmidrule(l){5-6}
                Platform & Method & Encode & Decode & Encode & Decode \\
                \midrule
                Pi~4 & SemCom (CLIP) & 661.1 & 0.5 & 647.4 & 0.5 \\
                Pi~4 & SemCom (MobileCLIP) & 889.9 & 0.5 & 657.7 & 0.5 \\
                Pi~4 & JPEG + 16-QAM & 1.4 & 674.5 & 3.0 & 661.7 \\
                Pi~5 & SemCom (CLIP) & 262.6 & 0.2 & 264.7 & 0.2 \\
                Pi~5 & SemCom (MobileCLIP) & 200.3 & 0.2 & 206.9 & 0.2 \\
                Pi~5 & JPEG + 16-QAM & 0.5 & 265.0 & 1.4 & 273.8 \\
                Pi~5+NPU & SemCom (CLIP) & 13.4 & 0.2 & 13.5 & 0.3 \\
                \bottomrule
            \end{tabular}
        \end{table}

    \subsection{Processing Time}
        \label{sec:experiment:runtime}
        Each scheme runs a single VLM forward pass per image, which dominates the computational cost, but the two schemes place it at opposite ends of the link (\autoref{tab:runtime}).
        In the semantic communication scheme the transmitter incurs it,
        and the receiver computes the cosine similarities as described in~\eqref{eq:decision},
        requiring only a single matrix-vector product with the cached prompt embeddings, which takes $0.497$\,ms on CIFAR-10 on the Pi~4.
        In the baseline scheme, the receiver incurs this cost,
        as it must decode the JPEG and re-embed the reconstructed image before making a decision, requiring $674.5$\,ms.
        Consequently, receive-side processing of zero-shot semantic communication is less costly, whereas transmit-side processing becomes more expensive.

        The mobile-scale model does not make encoding faster on the Raspberry Pi~4:
        although MobileCLIP is smaller, it encodes slower than CLIP on both datasets,
        because all Pi~4 inference runs in PyTorch on an unaccelerated Cortex-A72,
        whereas these architectures are designed to leverage mobile accelerators, \eg, the iPhone GPU.
        A Raspberry Pi~5 confirms this attribution: its Cortex-A76 cores encode $2.5\times$ faster for CLIP and up to $4.4\times$ faster for MobileCLIP,
        reversing the ordering so that MobileCLIP becomes $23.7\%$ faster on CIFAR-10 and $21.8\%$ faster on TSRD.
        A Hailo-10H NPU on the Pi~5 further reduces CLIP encoding to $13.4$\,ms, $19.6\times$ faster than its CPU and $49\times$ faster than the Pi~4,
        at the cost of $8$-bit quantization, which leaves accuracy unchanged at $83.3\%$ on TSRD and reduces it from $90.0\%$ to $88.0\%$ on CIFAR-10 without the radio link.
        Encoding therefore fits a real-time budget, so transmitter-side latency no longer limits the scheme.

        This asymmetry matters for practical deployment.
        In the baseline scheme a receiver aggregating multiple transmitters incurs the encoder cost once per link, whereas in the semantic scheme that cost is eliminated,
        so the receiver's load barely grows with the number of senders, \eg, drones, autonomous robots, or cybernetic avatars.

\section{Conclusion}
    \label{sec:conclusion}
    This study demonstrates the feasibility of zero-shot semantic communication on physical radio hardware.
    A Raspberry Pi equipped with two ADALM-Pluto transceivers successfully classifies previously unseen images using only the transmitted embedding.
    The implemented scheme needs $9\times$ fewer channel uses while achieving higher accuracy than the JPEG + 16-QAM baseline.
    The computational cost profile differs: encoding moves to the transmitter, but that is not a barrier,
    as a Hailo-10H NPU reduces it to $13.4$\,ms per image, $49\times$ faster than an unaccelerated Raspberry Pi~4 CPU, for at most two percentage points of accuracy.

    Several challenges remain.
    Because the present link uses a coaxial cable, repeating these measurements over the air in an anechoic chamber is the next step toward field deployment.
    Learned radio functions are also vulnerable to adversarial perturbations~\citep{le2025fggm}.
    Formal verification has been applied to deep learning-based traffic prediction~\citep{le2026formal}, massive MIMO power control~\citep{le2026formal2},
    and semantic communication under adversarial noise~\citep{le2026verifying}.
    Neural network verification~\citep{duong2025neuralsat,duong2026compositional,duong2026generating,duong2026clad,duong2026verifying3} now also covers vision~\citep{duong2026verifying,duong2026verifying2} and language~\citep{duong2026zonogpt} models,
    and extending it to the vision-language models used here could bound the receiver's error within the noise range measured on this testbed.
    Finally, because zero-shot semantic communication is agnostic to embedding content,
    it can carry embeddings from a vision-language-action model, instructing remote robotic agents with minimal data.

\section*{Acknowledgments}
This research is partially supported by JST Moonshot R\&D (JPMJMS2216).

\end{document}